\documentstyle[prl,aps,floats,epsf]{revtex}
\font\msbm=msbm9  scaled \magstep 1    
\def\kappa{\hbox{\msbm\char"7B}}       
\begin{document}
\draft
\tighten
\twocolumn[\hsize\textwidth\columnwidth\hsize\csname
@twocolumnfalse\endcsname
\title{ Onset of superconductivity and hysteresis in magnetic field for
a long cylinder\\
obtained from a self-consistent solutions of the Ginzburg--Landau equations}
\author{G.F. Zharkov}
\address{P.N. Lebedev Physical Institute, Russian Academy of Sciences,
 Moscow, 117924, Russia}
\date{21 March, 2001}
\maketitle
\begin{abstract}
Based on the self-consistent solution of a nonlinear system of
one-dimensional GL-equations, the onset and destruction of
superconductivity, the phase transitions and hysteresis phenomena
are discussed for a cylinder (radius $R$) in an axial magnetic field
($H$) for arbitrary $R,\kappa,H,m$ ($\kappa$ is the GL-parameter,
$m$ is the total vorticity of the system). The edge-suppressed
solutions (which are connected with the jumps of magnetization in
the states with fixed vorticity $m$), the depressed solutions
(responsible for the hysteresis in type-II superconductors), and
the precursor solutions (which describe the onset of
superconductivity in type-I superconductors) are also studied. The
limits of applicability of the so-called linear equation
approximation are discussed.
\end{abstract}
\vskip 2pc ] 
\narrowtext

\section{INTRODUCTION}

The macroscopic GL-theory [1] is widely used for studying the
behavior of superconductors in magnetic field. This theory leads
to two coupled nonlinear three-dimensional equations for a complex
order parameter $\Psi$ and magnetic field vector-potential ${\bf
A}$. Because, in a general case, it is impossible to find the
exact solution of this system analytically, the numerical
computations are used, or various limiting particular cases are
studied. Thus, the London theory [2] follows from the GL-equations
[1] in the limit $\psi\equiv |\Psi|=1$, and the Ginzburg approach
[3] corresponds to the case of a constant order parameter, which
depends on the weak external field $H$, $\psi=\psi(H)$. Another
approximation was formulated by Saint-James and de$\,$Gennes [4],
who assumed, that the nucleation of  superconductivity in large
fields is described by small values of the order parameter
($\psi\equiv|\Psi|\ll 1$). Instead of full system of nonlinear
GL-equations, they considered a single linear one-dimensional
equation for $\psi$, what allowed them to find analytically the
nucleation field $H_{c3}=1.69H_{c2}$ ($H_{c2}=\phi_0/(2\pi\xi^2)$,
$\xi$ is the coherence length) for a bulk superconducting slab in
parallel magnetic field. Later, Saint-James [5] used the same
approach to find the oscillation dependence of the nucleation
field $H_{c3}$ versus applied field $H$ for a very long circular
cylinder of finite radius $R$. In the papers [4,5] the basic
conception of the surface superconductivity was laid down,
according to which the onset of the superconductivity begins at
the fields $H\sim H_{c3}$ in a form of thin superconducting
sheath, layered near the surface of the specimen (the so-called
"giant-vortex state").

The structure of the giant-vortex solutions with finite values of
the order parameter ($\psi\sim 1$) was studied numerically in a
number of scattered publications (for the review of early works
see [6]). A more systematic study of such solutions, carried out
recently in [7], has revealed, that the axially symmetric solution
of fixed vorticity $m$ has different form, depending on the value
of the increasing external magnetic field $H$ ($m=0,1,2\cdots$ is
the quantum number, ensuring that $\Psi$ is single valued). In
type-II superconductors the giant-vortex solution has the
usual Meissner-type space-profile for fields $H<H_1$ (where
$H_1(m,R,\kappa)$ is some critical field). At $H=H_1$ this form
becomes unstable, and for $H>H_1$ the giant-vortex solution
acquires (by a first-order jump) a new "edge-suppressed" form [7].
With the field $H$ further increasing, the edge-suppressed
solution degenerates gradually and vanishes finally by a
second-order phase transition ($\psi\to 0$) at some field
$H_2(m,R,\kappa)$. (The field $H_2(m,R,\kappa)$ is just the
critical field, found in [5] from a linear theory; for $R\gg\xi$
and $m\sim 1$ the field $H_2$ coincides with $H_{c2}$; for
$R\gg\xi$ and $m\gg 1$ the maximum value of the field $H_2$
coincides with $H_{c3}=1.69H_{c2}$.)

In type-I superconductors the destruction of the superconducting state
occurs at the field $H_1(m,R,\kappa)$ (i.e. at the field of maximal
superheating)  by a first-order jump from $s$-state (having a finite
value of $\psi\sim 1$) to $n$-state ($\psi\equiv 0$). Because in this
case there are no solutions with $\psi\ll 1$, the linear theory is not
applicable, and the superheated $s$-boundary, found from the full
system of equations, deviates substantially from that, predicted in [5].

As is shown in the present paper, the nucleation of
superconductivity from $n$-state in decreasing fields also proceeds
differently in type-I and type-II superconducting cylinders.

In type-II cylinders the nucleation begins at $H_2(m,R,\kappa)$ as
a second-order phase transition from $n$-state, then a small
nucleated $s$-solution grows, repeating the edge-suppressed form
down to the field $H_1(m,R,\kappa)$. At this field the shape of
the order parameter starts to deviate gradually from the
edge-suppressed form and passes continuously into a new
"depressed" form, which exists down to some "restoration" field
$H_r(m,R,\kappa)$, where the first-order jump to the Meissner-type
solution occurs. In the field interval $\Delta=H_1-H_r$ there
exist simultaneously two solutions for the same field $H$: one is
the usual Meissner-type solution ($\psi\approx 1$), and the second
is the depressed solution ($\psi<1$), which describes the
hysteretic $s$-state of the cylinder. The region of parameters,
where $\Delta(m,R,\kappa)\ge 0$ (i.e. the phase boundary,where the
hysteretic transitions between superconducting states of the same
vorticity may exist), is found self-consistently.

In type-I cylinders (in the field decreasing regime) a supercooled
$n$-state persist down to some field $H_p$, where the feeble
($\psi\ll 1$) "precursor" solution forms, what corresponds to the
second-order ($n,s$)-phase transition, so the field $H_p$ may be
found from linear equation approximation [5]. However, the
precursor solution exists only in a very small interval
($\Delta_p$) of fields below $H_p$ (the field interval
$\Delta_p=H_p-H_r\sim 10^{-2} - 10^{-4} H_p$), after that the
first-order restoration of the full Meissner solution ($\psi\approx 1$)
occurs. [Thus, the nucleation of superconductivity in type-I
cylinders is an "almost first-order" phase transition.] The phase
region for type-I cylinders, where the magnetically "supercooled"
normal state [3] may exist, is also found self-consistently.

As was mentioned above, there are two different regimes to study
the effect of the external field penetration into the
superconductor interior: the field increase and the field decrease
regimes, which should be considered separately.  It is
demonstrated below, that if the field increases, there exists a
critical line $S_{\rm I-II}$ on the plane of variables
($m,R,\kappa$), which separates two superconducting regions,
$s_{\rm I}$ and $s_{\rm II}$. In the region $s_{\rm I}$ a
superconducting state vanishes by a first-order phase transition
to the normal state (by a jump from $\psi\ne 0$).  The
superconductors, which belong to the $s_{\rm I}$-region, is naturally
classified as type-I. (This classification is applied to
one-dimensional centrally symmetric states of a cylinder.) In the
region $s_{\rm II}$ (if the field increases) the destruction of
superconductivity is a second-order phase transition to normal
state ($\psi\to 0$). The superconductors, which belong to the
$s_{\rm II}$-region, naturally classified as type-II.

It is shown below (see also [7]), that the critical boundary
$S_{\rm I-II}$ between type-I and type-II superconducting
cylinders is not $\kappa_c=1/\sqrt{2}$ (as in the case of normal
and superconducting half-spaces, brought into contact [8]), but is
a complicated function of the cylinder radius $R$ and vorticity
$m$, i.e. $\kappa_c=f(R,m)$. It is shown, that for sufficiently
small $R$, all superconducting cylinders (independent of the
GL-parameter $\kappa$) belong to type-II class.

The paper is organized as follows. In Sec. II the basic
GL-equations are written, as well as the linear equation for
$\psi$. The known analytic solutions of the linear equation are
reproduced (in terms of the Kummer functions). The nonlinear
generalization of the Kummer-type equation is regarded. In Sec.
III the phase diagrams, which divide (on the plane of variables
$(R,\kappa)$) the regions of  first- and second-order phase
transitions to the normal state (and also the regions, where the
magnetic hysteresis is possible), are presented. The difference
between the depressed and edge-suppressed solutions is explained.

In Sec. IV the critical fields, found self-consistently from the
GL-equations for a cylinder in the increasing field, are compared
with those, found from the linear approach, for both type-I and
type-II superconductors. In Sec. V the analogous comparison is
made for the decreasing field. The "supercooling" of the normal
state, the precursor solutions and hysteresis in type-I
superconducting cylinders are also discussed. Sec. VI contains a
short resume and discussion of the results with possible
connection to experiment.

\section{EQUATIONS}

In what follows below, the case is considered of a long
superconducting cylinder of radius $R$, in the external magnetic
field $H\ge 0$, which is parallel to the cylinder element. Only
the radially symmetric (one-dimensional) solutions are studied. In
the cylindrical co-ordinates the system of GL-equations may be
written in dimensionless form [7]
$$
{ d^2U \over {dx^2} } - {1\over x}{ dU \over dx } - {\psi^2 U \over \kappa^2} =0,            \eqno(1)
$$
\vskip -0.5cm
$$
{d^2 \psi \over dx^2} + {1\over x}{d\psi\over dx} + (\psi - \psi^3) - { U^2  \over x^2}\psi =0.   \eqno(2)
$$
The dimensioned potential $A$, field $B$ and current $j_s$ are
related to the corresponding dimensionless quantities by the
formulae:
$$
A={ \phi_0 \over 2\pi\xi }{ U + m \over x }, \quad B={ \phi_0\over
2\pi\xi^2 }b, \quad b={1\over x}{dU \over dx},            \eqno(3)
$$
\vskip -0.5cm
$$
j(x)=j_s\Big/ { c\phi_0 \over 8\pi^2\xi^3 }= -\psi^2 {U\over x}, \quad  x= {r \over \xi}.
$$
The total vorticity $m$ in (3) specifies how many flux quanta are
associated with the vortex, centered at the cylinder axis (the
so-called giant-vortex state [5,9]). [The equivalent names for
$m$, are: vorticity, fluxoid, orbital momentum, magnetic quantum
number, winding quantum number.]

The boundary conditions to Eq. (1) are:
$$ U\big|_{x=0} = -m,\quad  \left. dU/dx\right|_{x =R_\xi}=h_\xi.  \eqno(4)
$$
where $R_\xi=R/\xi$, $h_\xi=H/H_\xi$, $H_\xi=\phi_0/(2\pi\xi^2)$.

The boundary conditions to Eq. (2) are:
$$
\left. d\psi/dx \right|_{x=0} =0, \quad \left.
d\psi/dx\right|_{x=R_\xi} =0 \quad (m=0),
$$
\vskip-1cm
$$
                        \eqno(5)
$$
\vskip-1cm
$$
\psi|_{x=0}=0, \quad \left. d\psi/dx \right|_{x=R_\xi} =0 \quad
(m>0).
$$

(We use here the coherence length $\xi$ as a unit of  length,
instead of the field penetration length $\lambda=\kappa\xi$, used
in [7]. Both length scales enter GL-equations on equal footing,
however, any of normalizations might be considered as preferable,
because the corresponding $\kappa$-dependencies in different
normalizations look sometimes differently.)

For small $\psi^2/\kappa^2\ll 1$ one can drop the last term  in
Eq. (1) [which accounts for the screening current],  so
$U=U_0(x)=-m+h_\xi x^2/2$, $B=H$ (no screening), and the system
(1),(2) reduces to a single equation
$$
{d^2 \psi \over dx^2} + {1\over x}{d\psi \over dx} + (\psi - \psi^3) - {U_0^2  \over x^2}\psi =0.       \eqno(6)
$$
We shall denote the solutions of the nonlinear equation (6) as
$\psi(x)=\widetilde{K}_m(x)$. The analytical form of the
(quasi-Kummer) functions $\widetilde{K}_m(x)$ is not known, but
they may be easily found numerically from Eq. (6) [see Sec. IV].

In the limit $\psi\to 0$, one can drop the term $\psi^3$ from Eq.
(6) and consider the linear equation [5]
$$
{d^2 \psi \over dx^2} + {1\over x}{d\psi \over dx} +  \left[1- {U_0^2  \over x^2} \right]\psi =0,         \eqno(7)
$$
with the solutions denoted as $\psi(x)=K_m(x)$. The function
$K_m(x)$ may be written in the analitical form [10]
$K_m(x)=y^{m/2}e^{-y/2}{\cal F}(y)$ (where $y=\gamma x^2$,
$\gamma=h_\xi/2$). The function ${\cal F}(y)$ satisfies the
confluent hypergeometric equation
$$
y{\cal F}''+(\mu-y){\cal F}'-\nu{\cal F}=0              \eqno(8)
$$
[where $\mu=m+1$ is a positive integer, $\nu=(1-h_\xi^{-1})/2$],
with a general solution ${\cal F}=C_1F+C_2\widetilde{F}$. The
function $F$ and can be written as an infinite series expansion
$$
F(\nu,\mu,y)=1+{\nu\over\mu}{y\over{1!}} + {\nu(\nu+1) \over
\mu(\mu+1)}{y^2\over {2!}}+\cdots.
$$
[We use the notation $F(\nu,\mu,y)$ [11--13] for the Kummer
functions, instead of $M(-n,l+1,y)$, used in [14,15]. If $\nu$ is
a negative integer, or zero, $F$ reduces to the polinom of power
$\nu$. The function $\widetilde{F}$ has a logarithmic singularity
at $x=0$ [16] and sometimes is dropped ($C_2=0$). However, the
function ${\cal F}(x)$ has a pre-factor $y^{m/2}\sim x^m$ (see
above), which cancels this singularity, so in a general case
$C_2\ne 0$ (for $m>0$).]

The second of the boundary conditions (5) is equivalent to
$$
\left. {d\over dy}K_m(\nu,\mu,y)\right|_{x=R_\xi}=0,    \eqno(9)
$$
which is a nonlinear equation for $\nu=(1-h_\xi^{-1})/2$  (if $m$
and $R_\xi$ are fixed). From Eq. (9) the proper value of $h_\xi$
(and also the space-profile $\psi(x)$) may be found. In a general
case, Eq. (9) has two roots for $h_\xi$ (if $m>0$), what
corresponds to two different solutions for $\psi(x)$ (see Sec.
IV). The maximum field $h_\xi$, above which there exists only the
solution $\psi(x)\equiv 0$, corresponds to the upper critical
field $h_2$.

[The solutions, found from Eqs. (1)--(9), describe the radially
symmetric giant-vortex states of fixed vorticity $m$. The more
complicated multi-vortex solutions (with the same total vorticity
$m=\Sigma_i m_i$, where $m_i$ is the vorticity of a vortex,
situated at some point $r_i$ on the cylinder cross-section) should
be described by the equations in partial derivatives and will not
be considered in the present paper. They may be studied
numerically by the methods, used, for instance, in [17--19] to
describe the results of experiments [20] with thin mesoscopic
discs.]

\section{THE PHASE VIEWS}

The solutions of Eqs. (1)--(5) depend on the space co-ordinate $x$
and several parameters [for instance,
$\psi(x)=\psi(m,R_\xi,\kappa,h_\xi;x)$; analogously for the
potential $U(x)$ and the field $b(x)$].  In this Section a general
view is given of the solutions structure in this many-parameter
nonlinear problem. Some details of this general picture will be
discussed in the ensuing Sections.

\subsection{Field increase regime}
\vskip -0.2cm

Let the vorticity $m$ be fixed ($m=0,\,1,\,2,\dots$) and consider
at first the case $m=0$ (the vortex-free Meissner state). Consider
the plane of parameters($R_\xi,\kappa$) (Fig. 1(a)). To every
point of this plane corresponds a set of solutions of
Eqs.(1)--(5), which depend parametrically on the external field
$h_\xi$ (or $h_\lambda=\kappa^2 h_\xi$). This set of solutions is
unique for each point ($R_\xi, \kappa$) and may be characterized,
for instance, by the field dependence of the maximum value of the
order parameter $\psi_{max}(h_\xi)$ in this point. It is helpful
to look at the plane $(R_\xi,\kappa)$ from above, and to imagine a
peep-hole, pierced in arbitrary point of this plane, which allows
to see the corresponding dependence $\psi_{max}(h_\xi)$. (Such
peep-holes are depicted as small circles in Fig. 1(a).) In doing
so, one can find three regions in Fig. 1(a), marked as $s_{\rm
I}^\Delta$, $s_{\rm II}^\Delta$ and $s_{\rm II}^0$. [The upper
index $\Delta$ denotes a possibility of the hysteresis in these
regions (see below). We shall mention Fig. 1(a), depicted on the
plane $(R_\xi,\kappa)$ (and similar ones, like those in Fig. 3),
as the phase views, to distinguish them from the phase diagrams,
depicted on the plane $(R_\xi,h_\xi)$ (as in Fig. 5).]

Fig. 1(b) shows (schematically) the behavior $\psi_{max}h_\xi)$ in
the region $s_{\rm II}^\Delta$ (the solid curve, the field $h_\xi$
increases from $h_\xi=0$). For small $h_\xi$, the field is almost
completely expelled from the superconductor (the Meissner state,
$m=0$, $b=0$). This part of the solid curve does not depend on
$h_\xi$ and may be described by the London theory [2]
approximation ($\psi=1$). In larger fields, the order parameter
depends on $h_\xi$ rather weakly. This part of the curve may be
described by the Ginzburg approximation [3]
($\psi=\psi(h_\xi)={\rm const}$). In still larger fields, a full
system of GL-equations is needed to describe the behavior
$\psi(h_\xi;x)$. There exists a critical field ($h_1$), at which
the stable Meissner-like solution (the upper section of the solid
curve) becomes absolutely unstable relative small perturbations in
its shape, and the solution $\psi(h_\xi;x)$ acquires new stable
space-profile, which may be called as the edge-suppressed state
[7] (see the lower part of the solid curve, which is labeled also
as "tail"; the difference between the various forms of solutions
is illustrated in Fig. 2). The transition from the upper to lower
solid branch in Fig. 1(b) is a first-order phase transition (with
a jump $\delta_1$). For fields $h_\xi>h_1$ the order parameter
diminishes gradually and vanishes finally at the field $h_2$ by a
second-order phase transition. Evidently, for fields $h_\xi\sim
h_2$ (when $\psi_{max}\ll 1$) the linear equation (7) is
applicable.

The width of the tail ($w=h_2-h_1$), where the superconducting
state ($m=0$) is destroyed by the second-order phase transition to
normal state, diminishes with the diminishing radius $R_\xi$. The
critical radius, at which the tail vanishes ($w=0$), depends on
$\kappa$, and is depicted in Fig. 1(a) by a solid line. This line
($S_{\rm I-II}$) divides the regions ($s_{\rm I}^\Delta$) of first
and ($s_{\rm II}^\Delta$) second-order phase transitions. In the
region $s_{\rm I}^\Delta$ (marked as "no tails"), the transition
to normal state is always of  first order, by a jump from a finite
value $\psi_{max}$ to $\psi\equiv 0$.

Type-I superconductors are naturally defined in our case as those,
which belong to the region $s_{\rm I}^\Delta$, and type-II -- to
the regions $s_{\rm II}^\Delta$, or $s_{\rm II}^0$. We see, that
the boundary between type-I and type-II superconductors is a
complicated function of $(R_\xi,\kappa)$, it depends on the
specimen geometry, and differs [7] from a simple phase boundary
$\kappa=1/\sqrt{2}$ [8], which describes the case of an infinite
(open) superconducting system. For instance, the superconducting
cylinder with $\kappa=1$ changes its type from type-II (above the
point $\alpha$ in Fig. 1(a)) to type-I (between the points
$\alpha$ and $\beta$), and again to type-II (below the point
$\beta$).

\subsection{Field decrease regime, type-II region}
\vskip -0.2cm

If the field $h_\xi$ decreases, starting from the large values
($h>h_2$, $\psi\equiv 0$), a weak superconducting state (with
$\psi_{max}\ll 1$), which nucleates at $h_2$, grows gradually into
the edge-suppressed form (with $\psi_{max}<1$ at $h_\xi=h_1$), but
for $h_\xi<h_1$ the edge-suppressed solution transforms
continuously into new "depressed" form (represented by the dotted
curve in Fig. 1(b)). In the field interval $\Delta=h_1-h_r$ there
exist simultaneously two stable solutions of Eqs. (1)--(5) (the
solid and dotted lines in Fig. 1(b)), the depressed solution being
responsible for the hysteresis in type-II cylinder. At the
restoration field $h_r$ the depressed solution becomes absolutely
unstable and transforms by a first-order jump ($\delta_r$) into
the Meissner-type form (a solid line; the space-profiles of
various solutions are illustrated in Fig. 2). The width of the
field interval, $\Delta=h_1-h_r$, where hysteresis is possible,
diminishes with the diminishing radius $R_\xi$. The critical
radius, at which $\Delta=0$, is presented by the dashed line in
Fig. 1(a). No hysteresis is possible below this line.

The first-order jumps ($\delta_r$ and $\delta_1$ in Fig. 1(b))
also diminish with radius, they disappear simultaneously at the
line $\Delta=0$ (Fig. 1(a)). At this line the two disconnected
pieces of the solid curve (Fig. 1(b)) met, and it acquires a
continuous form, with an inflection point (see Fig. 1(d)).  This
curve is reversible, in the sense, that for each $h_\xi$ there
exists only one solution, which is independent of the field
regime. For even smaller radius $R_\xi$ (in the region $s_{\rm
II}^0$, below the line $\Delta=0$ in Fig. 1(a)) the order
parameter $\psi_{max}(h_\xi)$ behaves as is shown by the dashed
line in Fig. 1(d). There is no hysteresis in the region $s_{\rm
II}^0$, and the transition to normal state is always of second
order, independently of $\kappa$-value.

\subsection{Type-I region}
\vskip -0.2cm

The order parameter $\psi_{max}(h_\xi)$ behaves very differently
in the type-I region $s_{\rm I}^\Delta$ of Fig. 1(a). If the field
increases, and if the radius $R_\xi$ is sufficiently large, and
the parameter $\kappa$ is sufficiently small, there are no tails
in $s_{\rm I}^\Delta$ (see Fig. 1(c)). The first-order jump
($\delta_1$) to the normal state occurs at the point $h_1$ (where
the superconducting solution in the field increase regime becomes
absolutely unstable). For fields $h_\xi>h_1$ the normal state
solution in Fig. 1(c) is absolutely stable, there are no other
solutions here, but $\psi\equiv 0$.

If the field $h_\xi$ decreases now below $h_1$, the supercooled
normal state solution ($\psi\equiv 0$) remains stable relative
small perturbations with $\psi\ne 0$, down to the point $h_p$. For
fields inside the interval $h_p<h_\xi<h_1$ there are two stable
solutions for $\psi$: one is $\psi\equiv 0$, the other ($\psi\sim
1$) is represented by the upper section of the solid curve in Fig.
1(c). (Both solutions are stable, because small perturbations in
there shape die down.) Thus, in the field interval
$\Delta=h_1-h_p$ the supercooling of the normal state is possible
[see the hysteresis loop in Fig. 1(c)].

Below $h_p$ the normal state becomes absolutely unstable (small
perturbations grow), and a feeble superconducting solution (or,
the "precursor" state, $\psi\ll 1$) establishes in the bulk of a
cylinder. (The position of $h_p$ coincides exactly with the
critical field of  $\kappa$-independent second-order
($s,n$)-transitions and may be found from a linear theory [5].)
The precursor solution exists down to some restoration field $h_r$,
where it becomes absolutely unstable and the Meissner form restores
[notice the presence of first-order jump $\delta_r$ in Fig. 1(c)]. The
field interval $\Delta_p=h_p-h_r$, where the precursor solution
exists, is usually very small ($\Delta_p\sim 10^{-2} - 10^{-4} h_p$),
so in type-I superconductors the restoration of superconductivity
from a supercooled $n$-state is "almost first-order" phase transition.
The exact value of the restoration field ($h_r$) and the
corresponding amplitude of the precursor solution ($\psi_r$) can
not be found from the linear theory, but self-consistent solution
of the full system of nonlinear GL-equations is required.

If the radius $R_\xi$ diminishes, the jumps $\delta_r$ and
$\delta_1$ also diminish (as well as the interval
$\Delta=h_1-h_p$, see the dashed curves in Fig. 1(c)), they all
disappear simultaneously with $\Delta\to 0$. The line $\Delta=0$
(for small $\kappa$ and $R_\xi$) merges with the line $S_{\rm
I-II}$ in Fig. 1(a) at the point $\beta$ (compare the dashed
curves in Figs. 1(c) and 1(d)). No supercooling of normal state is
possible below the line $\Delta=0$.

\subsection{Examples of co-ordinate dependencies}
\vskip -0.2cm

The co-ordinate dependence of the solutions [as they are seen
through the peep-hole, pierced in the type-II point $R_\xi=10,
\kappa=1.2$ in Fig. 1(a)] is shown in Fig. 2 for different values
of the external field $H$ ($h_\xi= H/H_\xi$). In the field
increasing regime, the sequence of curves, which result from Eqs.
(1)--(5), are numerated as ({\it 0, 1, 2, 2e,} $K_m$). The order
parameter $\psi(r)$ (Fig. 2(a)) changes its form continuously from
{\it 0} ($\psi\equiv 1$ at $h_\xi=0$) to {\it 1} ($h_\xi=0.9037$),
and then to {\it 2} ($h_\xi=1.0951$ -- this is the critical field,
marked as $h_1$ in Fig. 1(b)). At this point the solution {\it 2}
becomes absolutely unstable and (by a first-order jump,
$\delta_1=0.155$) acquires the edge-suppressed form {\it 2e}
($h_\xi=1.0952$). If the field $h_\xi$ increases further, the
solution {\it 2e} passes continuously into the form $K_m$ (the
Kummer-type solution), which vanishes finally at the critical
field $h_\xi=h_2=1.0016$, where $\psi\equiv 0$. Fig. 2(b)
illustrates the behavior of the induction $B(r)$ versus the
external magnetic field $H$. At the point $h_1$ the solution {\it
2} switches (by a jump $\delta_1$) from the Meissner form {\it 2}
(the external field is effectively screened out) to the
edge-suppressed form {\it 2e} (the external field penetrates the
edge region without screening). The curve $K_m$ may be described
by the linear equation approximation ($B=H$).

The sequence of solutions, appearing in Fig. 2 in the field
decreasing regime, is different (see Fig. 1(b)). The normal state
solution ($\psi\equiv 0$) is absolutely stable for $h_\xi>h_2$,
and absolutely unstable for $h_\xi<h_2$. If the field decreases, a
small superconducting solution (of the Kummer-type, $K_m$) appears
at $h_\xi=h_2$ and transforms continuously into the
edge-suppressed form {\it 2e} (with $\psi\sim 1$ at $h_\xi=h_2$).
If the field $h_\xi$ decreases below $h_1$, the solution {\it 2e}
keeps transforming continuously into the form {\it 1d} (at
$h_\xi=h_r$). The intermediate sequence of curves ({\it 2e}
$\iff${\it 1d}) presents the depressed solutions, which are
characterized by a smaller average values of the order parameter
$\overline{\psi}$, in comparison with the Meissner solutions ({\it
1}$\iff${\it 2}). For $h_\xi<h_r$ the depressed solution {\it 1d}
becomes absolutely unstable and transforms (by a jump $\delta_r$
at $h_\xi=h_r$) into the Meissner form {\it 1}. It is important,
that in the interval of fields $\Delta=h_1-h_r$ there exist
simultaneously (for the same $h_\xi$) two independent solutions of
Eqs. (1)--(5) (the Meissner and depressed forms). The presence of
two solutions means a possibility of hysteresis in the system.

[The Meissner solution can be obtained from Eqs. (1)--(5) by the
iteration procedure [21], started with a large trial function
$\overline{\psi}\sim 1$. The depressed form is obtained by the
iterations, started with a small trial function
$\overline{\psi}\sim 0.01$. Outside the interval $\Delta=h_1-h_r$
both iteration procedures produce the same self-consistent
result.]

Additional examples of co-ordinate dependencies of the solutions
for various $R$ and $\kappa$ may be found in [7].  (The
space-profiles of the  precursor solutions are depicted below in
Fig. 8.)

\subsection{The phase views for m=1,\,2}
\vskip -0.2cm

The analogous phase views on the plane $(R_\xi,\kappa)$ for the
vortex states $m=1$ and $m=2$ are depicted in Figs. 3(a,b), they
are similar to the phase view $m=0$ in Fig. 1(a). The only
essential difference is in the presence of the minimal radius
$R_\xi$ [see the dashed lines $C_{sn}$ in Figs. 3(a,b)], below
which only the normal state solutions are possible. This is
natural, because in a case of very small radius specimens the
intrinsic magnetic field of a vortex is too strong for the
survival of the superconducting state.

Fig. 4 illustrates the solutions behavior in the vortex states
$m=1,\, 2$, as they are seen through the peep-holes in type-II
points $R_\xi=10$, $\kappa=1.2$ in Figs. 2(a,b). If the field
increases, the sequence of the soluions is ({\it 0, 2, 2e},
$K_m$). In the Meissner-type states ({\it 0, 1, 2}) the external
field is screened out. At the field $h_\xi=h_1$ the jump
transformation to the edge-suppressed form ({\it 2}$\to${\it 2e})
occurs (with a jump $\delta_1$). (The concrete values
$h_r,h_1,h_2,\delta_r,\delta_1$ are given in the caption to Fig.
4.)

In the field decreasing regime the sequence of the appearing
solutions is ($K_m$, {\it 2e, 1d, 1, 0}), with a jump
transformation ($\delta_r$) from {\it 1d} to {\it 1}. In the field
interval $\Delta=h_1-h_r$ there are two solutions: one is the
Meissner-type form (intermediate between {\it 1} and {\it 2}), and
the other is the depressed form (intermediate between {\it 2e} and
{\it 1d}). Within the field interval $\Delta$ the hysteresis
transitions between the Meissner-type and depressed states of the
same vorticity $m$ are possible. For fields outside the interval
$\Delta$ no hysteresis is possible, because only one solution
exist there (see Fig. 1(b)): either the Meissner-type solution
(for $h_\xi<h_r$), or the edge-suppressed solution (for
$h_\xi>h_1$).

The giant vortex states with $m>2$ may be studied analogously.

\section{THE PHASE DIAGRAMS}

In this Section the comparison is made of the phase diagrams (or,
the critical fields $h_2$) for the second-order
($s,n$)-transitions in the increasing magnetic field, found
self-consistently and in the linear approximation. [See also Ref.
[7a], where the critical fields $h_1$ for the first-order jumps to
the edge-suppressed states were studied.]

From the phase view of Fig. 1(a) it follows, that if the magnetic
field increases, the order parameter vanishes either gradually
[with a tail in the curves of Figs. 1(b,d)], or by the first-order
jump [Fig. 1(c)]. Fig. 5(a) is a phase diagram, which shows the
dependence of the critical field $h_{sn}^c$ in Fig. 1 [i.e. the
field $h_1$  (type-I), or $h_2$ (type-II), for which the
transition from $s$- to $n$-state occurs (in the field increase
regime)], as a function of $R_\xi$, for different $\kappa$ and
$m=0$. The thick line $K_m$ corresponds to the $(s,n)$-diagram,
found from the linear equation (7) with the boundary conditions
(5).

As is seen from Fig. 5(a), for $\kappa>1$ the self-consistent
critical curves $h_{sn}^c(h_\xi)$ coincide with the universal
curve $K_m(h_\xi)$ (found from the linear equation). Thus, for
type-II superconductors (with $\kappa>1$) the linear approach
[4,5] gives correct description of the $(s,n)$-boundary.

However, for type-I superconductors (with $\kappa<1$), the
critical curves $h_{sn}^c(h_\xi)$ deviate from the universal curve
$K_m(h_\xi)$ considerably. The inspection of the asymptotes of the
curves $h_{sn}^c$ (for $R_\xi\gg 1$) shows, that they have
different functional dependences versus parameter $\kappa$. For
type-II superconductors ($\kappa>1$) the asymptotes are
$h_{sn}^c\approx 1$ (they are $\kappa$-independent, in accordance
with the linear theory [5]). For type-I superconductors
($\kappa<1$) the asymptotes behave as $h_{sn}^c\approx
\kappa^{-p}$ ($1<p<2$); this means, that the universal linear
theory fails for small $\kappa<1$ and large $R_\xi$. (For
sufficiently small radius $R_\xi$ the ($s,n$)-transition is always
of second order, so in this case the linear theory is valid for
all $\kappa$.)

The analogous conclusions may be drawn from Fig. 5(b) ($m=1$) and
Fig. 5(c) ($m=2$).

The fact, that the critical fields $h^c_{sn}$ in Fig. 5 for
$\kappa<1$ deviate from the predictions of the linear theory [5],
can be attributed to strong nonlinear competition between two
characteristic lengths $\lambda$ and $\xi$ in GL-equations. For
$\kappa=\lambda/\xi>1$ the nonlinear equations can be linearized,
but for $\kappa=\lambda/\xi<1$ they can not. (Evidently, the
limitations for the linear theory can not be deduced from the
linear theory itself, but self-consistent analysis of full
nonlinear system of equations is required.)

\subsection{Two solutions of the linear equation}
\vskip -0.2cm

Notice the peculiar behavior of the phase diagrams in Figs.
5(b,c). For sufficiently small $R_\xi$ the line $R_\xi=$const
crosses the bottom part of the curves $K_m$ at two points. This
means the existence of two solutions $K_m(x)$ of the linear
equation (7), which both satisfy the boundary conditions (5) for
the same $R_\xi$, but with different $h_\xi$ (see Eq. (9)). These
two solutions are depicted in Fig. 6(a) (the solid lines {\it 1}
and {\it 2}), as they are seen through the peep-holes in Fig.
5(b), pierced along the line $R_\xi=1.5$ at the points
$h_\xi=0.67$ and $h_\xi=2.32$ ($m=1$).

[The insert to Fig. 6(a) illustrates (schematically) the behavior
of the solutions of the hypergeometric equation (7) for $m=1$. The
function $K_m(x)$, in a general case, has two extremal points $A$
and $B$, where $K_m'(x)=0$ (marked by the arrows), whose positions
depend on $h_\xi$. The curves $A$ and $B$ in Fig. 6(a) have
different functional behavior: the curve $A$ grows monotonously,
reaching the extremum at $x=R_\xi=1.5$. The curve $B$ has a zigzag
in the vicinity of $x=R_\xi$ (the zigzag amplitude $\delta$ is
small for small $R_\xi$ and is not seen in this scale). The
functions $K_m(x)$ for $m>1$ behave analogously, see Fig. 6(b) for
$m=2$, $R_\xi=2.5$.]

The existence of two different solutions of Eq. (7) has clear
physical interpretation. In the absence of the external field, the
vortex is held inside the cylinder by pinning to the boundary
(which is the source of the inhomogeneity in otherwise homogeneous
system). However, in small radius cylinders the internal vortex
field $B_i$ is too strong to be confined inside the mesoscopic
sample by the pinning force, and it breaks outside. To prevent the
field $B_i$ from leaking to the outer space, it is necessary to
impose a finite external field $H$, which helps to keep $B_i$
inside the specimen. (This can be considered as an example of the
so-called re-entrant superconductivity, or, the field-enhancement
effect.)\, However, if the field $H$ increases, the
superconductivity will be finally destroyed. Two solutions ($A$
and $B$ in Figs. 6(a,b)) describe two different physical
situations. The solution $B$ corresponds to the superconducting
vortex state ($m>0$) being destroyed by a large external field
$H$. The solution $A$ (with smaller $H$) corresponds to the vortex
state being destroyed by the internal field of it's own vortex.

The solutions $A$ and $B$ in Figs. 6(a,b) are depicted as they are
seen through the peep-holes in Figs. 5(b,c), which lie near the
opposite sides of the thick curves $K_m$ [these curves represent
the states with $\psi_{max}\ll 1$; the corresponding solutions
$\psi(x)$ are the hypergeometric functions $K_m(x)$ of the linear
Eq. (7)]. In the intermediate peep-holes $i$ (Fig. 5) the
solutions $\psi(x)$ have finite amplitude $\psi_{max}$, but (if
$\psi_{max}$ is yet sufficiently small) the solution may be
approximated by the function $\widetilde{K}_m(x)$ of the nonlinear
Eq. (6) [see the dashed curves $C$ in Figs. 6(a,b)]. To find
$\psi(x)$ with still larger $\psi_{max}$, it is necessary to solve
the full system of Eqs. (1),(2). Such solution is presented by the
curve $D$ in Fig. 6(b).

\subsection{Supercooling of the normal state}
\vskip -0.2cm

As was explained in Sec.III, in any peep-hole in the region
$s_{\rm II}^\Delta$ of Fig.1(a) (where$\kappa>1$ and $\Delta>0$)
there exist two solutions of Eqs. (1),(2): one is the
Meissner-type solution, and the other is the depressed (or
partially suppressed) solution, which is responsible for the
hysteresis in type-II superconductors (see Fig. 2 for $m=0$ and
Fig. 4 for $m=1,\,2$).

The analogous hysteresis phenomena exist also in type-I
superconductors [in the region $s_{\rm I}^\Delta$ of Fig. 1(a),
where $\kappa<1$ and $\Delta>0$; analogously for Fig. 3]. In this
case, there are also two solutions: one is again of the Meissner
type, but the partially depressed solution (represented by the
dotted curve in Fig. 1(b)) degenerates now into the totally
depressed normal state of Fig. 1(c). Thus, in type-I
superconductors the second branch of solutions in Fig.1(b)
corresponds to the supercooled $n$-state ($\psi(x)\equiv 0$).
(Notice, that in type-II superconductors the supercooled normal
state can not exist.)

The curve $sh$ in Fig. 7(a) corresponds to a maximal field, at
which the superheated Meissner state ($m=0$) may still exist in
type-I superconductor with $\kappa=0.1$ in the field increase
regime (i.e. the field $h_2$ in Fig. 1(c)). The curve $sc$
corresponds to a minimal field, at which the supercooled normal
state may still exist in the field decrease regime (i.e. the field
$h_r$ in Fig. 1(c)). These two curves merge at the point LCP (the
Landau critical point [22]) into a single curve (the calculated
difference between $sc$- and $sh$-curves at LCP is less than
$1\cdot 10^{-4}$). The free energies of the superconducting
states, which lie between $sc$- and $sh$-curves, depend on the
field $h_\xi$, and at some value of $h_\xi$ the difference of free
energies $\Delta G=G_s-G_n$ vanishes. The curve $eq$ in Fig. 7(a)
shows (schematically) the position of the equilibrium curve
($\Delta G=0$). All three curves merge at LCP, which is a
three-critical point.

The analogous dependences are depicted in Fig. 7(b) for
$\kappa=0.5$. It is clear, that the width of the region between
$sc$- and $sh$-curves diminishes with $\kappa$ increasing; for
$\kappa=1$ both curves merge into a single curve and the
(normal-state) hysteresis region disappears.

Notice, that for $R_\xi\gg 1$ the maximal supercooling of the
normal state in type-I superconducting cylinder is reached in the
field $h_\xi=1$, i.e. in the field
$H_\xi=H_{c2}\equiv\phi_0/(2\pi\xi^2)$. To the same conclusion
came Ginzburg [3] and Abrikosov [8], who used the approximations
$\psi={\rm const}$ and $\kappa\ll 1$. [The dependence of the
superheating field $H_{sh}(H)$ for type-I cylinders was found
earlier (in a different parametrization) by Esfandiary and Fink
[23], who used self-consistent solutions of the nonlinear
equations (the supercooling  field $H_{sc}$ was not studied in
[23]). (See [6] for the review of early theoretical and
experimental works on the problem of hysteresis in type-I
superconductors.)]

The comparison of $sc$-curves in Figs. 7(a,b) with the universal
$K_m$-curve in Fig. 5(a) reveals, that they are identical and have
the following asymptotic behavior: $h_\xi(R_\xi)=1$ for $R_\xi\gg
1$; $R_\xi(h_\xi)=2.8/h_\xi$ for $h_\xi>2$ (see dotted lines
$\alpha$ in Figs. 7(a,b)). Notice, that these curves coincide with
each other for all $R_\xi$ and $\kappa< 1$ (not only for
$R_\xi\gg 1$ and $\kappa\ll 1$, as was found in [3,8]). This
coincidence is not accidental, because at the supercooling
$n$-boundary the precursor solutions appear ($\psi_{max}\ll 1$)
which are described by the $\kappa$-independent linear equation
(7).

The precursor solutions are depicted in Fig. 8 for the fields
$h_\xi$, which just precede the jumps to the Meissner state with
$\psi\sim 1$. The precursor solution shape is $\kappa$-independent
(in accordance with Eq. (7)), but the exact position of the jump
and the amplitude of $\psi_{max}$ are $\kappa$-dependent and
should be found from the full system of GL-equations (1), (2).
Notice, that the precursor solutions describe the
superconductivity nucleation in supercooled type-I cylinder as a
bulk (though very feeble) effect. [The detailed study of the
precursor solutions behavior would be presented elsewhere.]

\subsection{The free-energy functional}
\vskip -0.2cm

As was mentioned above, the problem of supercooling and
superheating in type-I superconductors was considered first by
Ginzburg [3], who analysed the behavior of the free energy
functional in the approximation $\psi={\rm const}$ ($m=0$). It is
expedient to compare the results of the self-consistent and
approximate approaches.

In Fig. 9(a) the exact dependence $\psi_{max}(h_\lambda)$ is shown
for $\kappa=0.5$ at $R_\xi=3$ and $R_\xi=1$ ($m=0$). It is
evident, that the width $\Delta$ of the hysteresis region
diminishes with $R_\xi$ diminishing, and vanishes at some
$R_{min}$ (at the point LCP in Fig. 7(b)).

In Figs. 9(c) the exact field dependence is shown of the
normalized free energy ($\Delta g$) for $\kappa=0.5$ and
$R_\xi=3$. (The expressions for $\Delta g$ and magnetization
($-4\pi M$) may be found in [7]). Evidently,  the part of the
curve $\Delta g(h_\xi)$, which lies to the right of the point $eq$
(where $\Delta g=0$), corresponds to the metastable $s$-state with
$\Delta g>0$. In this region $s$-state is energetically unstable
(because $n$-state has smaller free energy, $\Delta g=0$).
[However, the energetically unstable $s$-state is stable relative
small spatial perturbations.] To the left of the point $eq$ the
$s$-state is energetically more favorable ($\Delta g<0$), and the
$n$-state (with $\Delta g=0$) is metastable. [However, the
energetically metastable $n$-state is stable relative small
spatial disturbances.] At the equilibrium point $eq$ the
transition from one branch of the solution to the other (with
smaller free energy) may happen. In this case the system behavior
would be totally reversible (without hysteresis).

In Fig. 9(b) the field dependence of the magnetization ($-4\pi M$)
is illustrated. In the case of equilibrium transition the
reversible jump of the magnetization should be observed at the
point $eq$. If the metastable states are realized, the jumps of
the magnetization may happen anywhere between the points $sc$ and
$sh$, with the accompanying hysteresis. [There is a principal
difference between the jump transitions in Fig. 1(b) and Fig. 8.
The first-order jumps in type-II superconductors (Fig. 1(b)) occur
within the same state ($m=0$), while the jump transitions in
type-I superconductors (as in Fig. 9) occur between different
states ($s$ and $n$).]

Notice, that in the approximation $\psi={\rm const}$ [3] the free
energy $\Delta G(\psi)$ has the form, schematically depicted in
Fig. 10(a) (the curves {\it 1}$\div${\it 5} correspond to the
increasing fields $H$). The extremas of these curves define the
possible values of the order parameter $\psi_0(H)$, which are
depicted (schematically) in Fig. 10(b). [The numerals {\it
1}$\div${\it 5} in this figure mark the positions of the extremas
on the corresponding curves in Fig. 10(a).] Thus, in the
approximate approach [3] there are two solutions for $\psi_0\ne 0$
in the field interval $H_{sc}<H<H_{sh}$, one (which corresponds to
the minimum of the free energy $\Delta G$) is energetically
stable, the other (which corresponds to the maximum of the free
energy) is energetically unstable.

However, according to the self-consistent theory, in type-I
superconductors there exist only one solution with $\psi\ne 0$
[which might be energetically stable, or metastable, depending on
$H$ (see Fig. 9(a))], the other (with $\psi\equiv 0$) corresponds
to the supercooled normal state. Thus, according to the exact
theory, the unstable branch $\psi_0\ne 0$ of approximate theory
does not exist at all.

This contradiction arises because of the following reason. In the
rigorous theory the free-energy is a functional, $G[\psi(x)]$,
which produces a number, $G$, from a function, $\psi(x)$. The
function $\psi(x)$ depends parametrically on $H$, so the
functional $G[\psi(x)]$ is a function of $H$. In the Ginzburg
approximation the functional $G[\psi(x)]$ is replaced by a
function $G(\psi_0,H)$, where $\psi_0$ is considered as an
arbitrary number. This function has minima and maxima, what is
illustrated by Fig. 10. However, as follows from the
self-consistent solution of nonlinear problem, some of the values
$\psi_0$ are forbidden. As can be seen from Fig. 9(c), the exact
functional $G(H)$ increases with $H$, but it does not reach an
extremal point, where $G+(H)=0$, and drops to zero at some value
of $H$, where $\psi(H)$ terminates. Thus, the Ginzburg
approximation for $G(\psi_0)$ [3] is qualitatively valid, if
$\psi(x)\approx\psi_0={\rm const}$ [see the stable (solid) branch
of the curve $\psi_0(H)$ in Fig. 10(b)], but fails for those
(forbidden) values of $\psi_0$, which belong to the unstable
branch of $\psi_0$ (the dashed curve). Nevertheless, the point of
maximal supercooling,  $H_{sc}$, is described by the Ginzburg
approximation correctly, because at this point ($\psi_0=0$) the
precursor state nucleates. [The precursor solution has very small
amplitude, $\psi(x)\approx 0$, what always satisfies the condition
[3] $\psi(x)\approx{\rm const}$, independently of the real
space-profile of $\psi(x)$ (see Fig. 8).]

We shall restrict ourselves by these methodical remarks, while
comparing the results of the rigorous and approximate approaches.

\section{CONCLUSIONS AND DISCUSSION}

We have studied in detail the one-dimensional solutions of
GL-equations in a case of cylinder geometry. The main new results
of the present investigation are summarized below. Mention among
them: the phase views, presented in Figs. 1(a) and 3(a,b); the
existence of the edge-suppressed and depressed solutions
(Figs.$\!$ 1,\,2,\,4); the phase diagrams of Fig. 5; the
discussion of the applicability of the linear equation
approximation (Sec.$\!$ V); the discussion of two independent
solutions of the linear equation (Fig. 6); somewhat novel insight
into the problem of hysteresis in type-I superconductors (Figs.
7--9) and the discussion of the previously unknown precursor
solutions; the discussion of the free-energy functional and the
comparison of the rigorous and approximate approaches to the
hysteresis problem (Fig. 10).

In conclusion, some additional comments to the topics, discussed
above, should be made.

The edge-suppressed solutions in a cylinder geometry were
described first by Fink and Presson in a footnote to their paper
[9], but were disregarded as being unstable in the energy space.
We consider these solutions (as well as depressed and precursor
solutions) as stable in the co-ordinate space. Mathematically,
they are usual axially symmetric solutions, which have
different forms in different regions of parameters (due to the
nonlinearity of the problem). These solutions describe symmetrical
$m$-states, which the physical system may occupy. At the critical
field $h_1$ the Meissner state ($m=0$) becomes unstable and the
external field fluds the outer region of a cylinder (see the
edge-suppressed solution {\it 2e} in Fig. 2). Another possible
mechanism for the field penetration is a creation of a vortex line
($m=1$) on the cylinder axis (see solutions in Fig. 4). Comparing
the free-energies of the competing states, one can find the points
of equilibrium transitions [7], and also the field interval, where
the edge-suppressed state may exist as a metastable one. The
metastable states may manifest themselves in experiments, where
such metastability is realized (for instance, in the hysteresis
phenomena [24], in paramagnetic Meissner effect [25], etc.),
especially in a case of mesoscopic samples (with few number of
vortices), when the stabilizing role of the boundary is important.

It is interesting, that similar effects -- such as jumps of the
magnetization within the states of fixed $m$, the transition to
the edge-suppressed form of a giant-vortex solution, the
re-entrance of superconductivity and the field-enhancement effects
-- exist also in a case of mesoscopic disks of small height,
studied self-consistently in [17,18]. This means, that these
nonlinear effects are common to various sample geometries and are
determined mainly by the parameters, entering the GL-equations,
so, the role of the boundary is, probably, not crucial.

No immediate comparison between the present theory and experiment
was attempted, partly because the model case of infinitely long
cylinder approximates only remotely the thin-disk geometry, mostly
used in recent experiments [20]. Moreover, in discussing such
experiments, it is necessary to consider also the asymmetric
multi-vortex solutions (see the end of Sec. II), what requires
using specific numerical methods and large computers (see, for
instance, [17,18]). However, a number of qualitative predictions,
obtained from the one-dimensional theory, may be used to interpret
some of the peculiarities, observed on mesoscopic samples. For
instance, it might be possible to attribute some jumps of the
magnetization (seen in [20,24,25]) not to transitions between
different $m$-states, but occurring within the same $m$-state (due
to the order-parameter reconstruction, while it passes to the
edge-suppressed form). However, a special experimental analysis
would be needed to confirm the existence of such transitions.

Notice, that the precursor solutions, studied above (see Fig. 8),
show, that  type-I superconductor may be found in two different
(metastable) hysteretic states (in addition to the stable Meissner
state). One is a supercooled normal metal ($\psi\equiv 0$); this
state may persist down to the maximal supercooling field ($h_p$).
The other is the supercooled precursor state ($\psi\ne 0$), which
nucleates at $h_p$ and survives down to the restoration field
($h_r$), when the first-order transition to the Meissner state
($\psi\approx 1$) occurs. This precursor state is more prominent
for values of $\kappa$, laying in the vicinity of $S_{\rm
I-II}$-phase boundary (Fig. 1(a)). Though the field interval
$\Delta_p$, where the precursor state may be found, is small, it
would also be of interest to seek for experimental evidences of
its possible existence.

Evidently, further theoretical and experimental study of the
problems, discussed above, is desirable.

[The one-dimensional solutions in other geometries (a plate of
finite thickness and superconducting half-space) were addressed
earlier in a number of publications (see, for instance, [26--30]).
This topic will be discussed elsewhere.]

\section{Acknowledgments}

I am grateful to V.L.Ginzburg for the interest in this work and
illuminating discussions. The valuable discussions with F.Peeters,
S.Yampolskii and B.Baelus (University of Antwerpen) are also
acknowledged with gratitude. I thank prof. R.Gross for hospitality
and helpful discussions during my stay at the
Walther-Meissner-Institute (Garching), where the final version of
this paper was written. The initial stage of this work was partially
supported through the grant N00173-99-P-0186.

\centerline{\bf Figures captions}

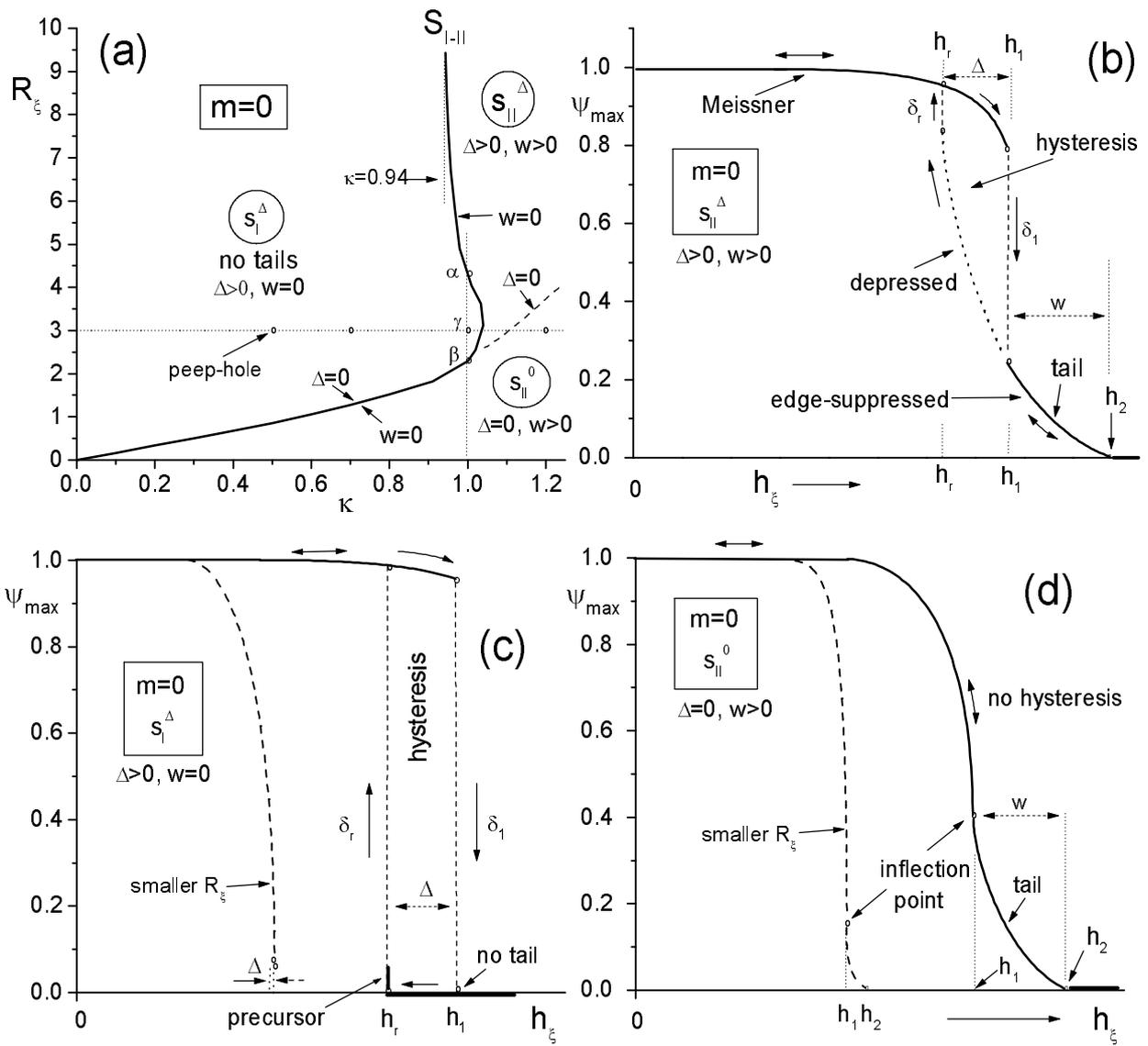
Fig. 1. (a) -- The phase view for a cylinder in the vortex-free
Meissner state $(m=0)$ in the field increasing regime; $s_{\rm
I}^\Delta$ -- the region of the first-order phase transitions from
$s$- to $n$-state; $s_{\rm II}^\Delta$  and $s_{\rm II}^0$ -- the
regions of second-order phase transitions. The curve $S_{\rm
I-II}$ divides the regions of first- and second-order phase
transitions. In the region above the dashed line $\Delta=0$ the
hysteresis transitions are possible; no hysteresis is possible
below this line. (b) -- Schematic behavior of the order parameter
$\psi_{max}(h_\xi)$ in the region $s_{\rm II}^\Delta$ ($m=0$,
$\Delta>0$, $w>0$). (c) -- Schematic behavior of the order
parameter in the region $s_{\rm I}^\Delta$ ($m=0$, $\Delta>0$,
$w=0$). (d) -- Schematic behavior of the order parameter in the
region $s_{\rm II}^0$ ($m=0$, $\Delta=0$, $w>0$). The used
notations are explained in the text.

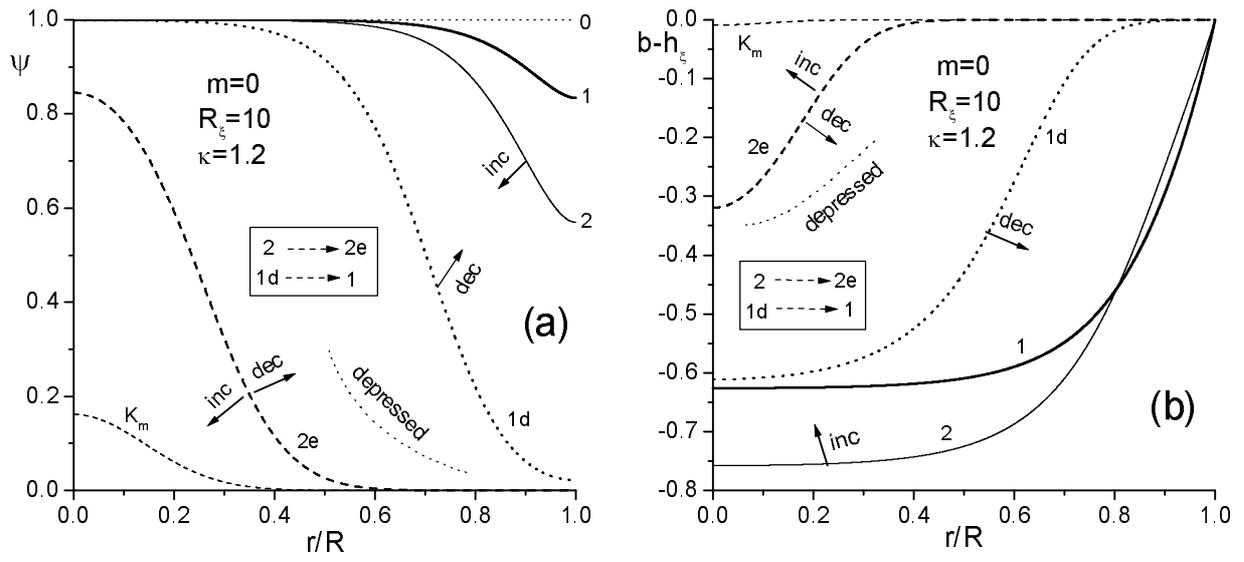
Fig. 2. The coordinate dependences: (a) -- for $\psi(x)$ and (b)
-- for $b(x)$, as they are seen through the peep-hole in Fig. 1(a)
at $R_\xi=10$, $\kappa=1.2$ ($m=0$). The curves are calculated for
the dimensionless fields $h_\xi=H/H_\xi$: {\it 0} -- $h_\xi=0$;
{\it 1} -- $h_\xi=0.6275$; {\it 1d} -- $h_\xi=0.6276$; {\it 2} --
$h_\xi=0.7605$; {\it 2e} -- $h_\xi=0.7606$; $K_m$ --
$h_\xi=0.992$; $\psi_{max}=0$ at $h_\xi=1.000$. The jump
$\delta_1$ of $\psi_{max}$ (see Fig. 1(b)) at the transition {\it
1d}$\to${\it 1} is $\delta_1=2\cdot 10^{-4}$. The jump at the
transition {\it 2}$\to${\it 2e} is $\delta_r=0.155$. The arrows
show how the solutions transform, if the field $h_\xi$ is
increased, or decreased. The depressed solutions, responsible for
the hysteresis, have the form, intermediate between {\it 2e} and
{\it 1d}.

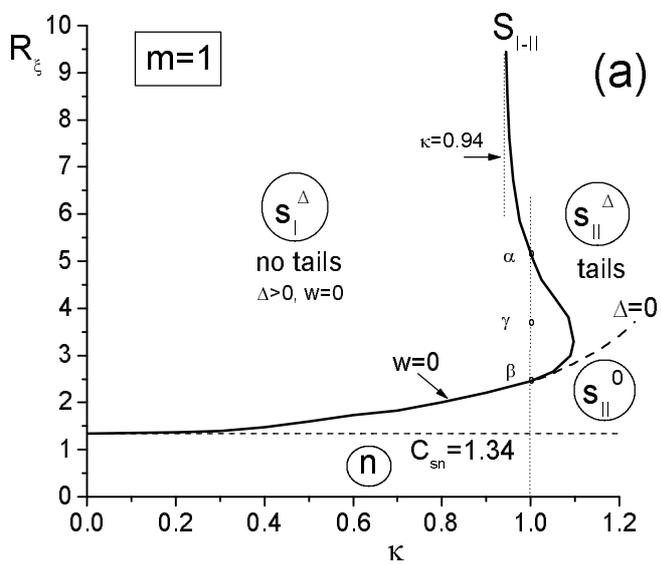
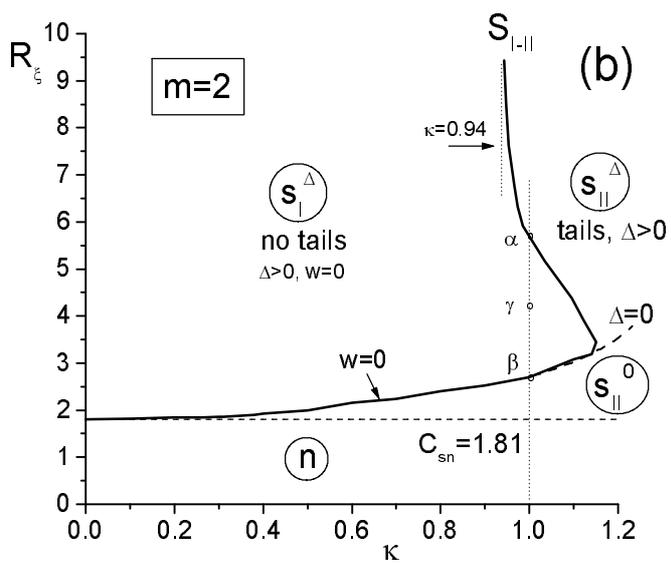
Fig. 3. The phase views: (a) -- for $m=1$ and (b) -- for $m=2$.
Notations are the same, as in Fig. 1(a). Below the line $C_{sn}$
lies the normal-metal region $n$.

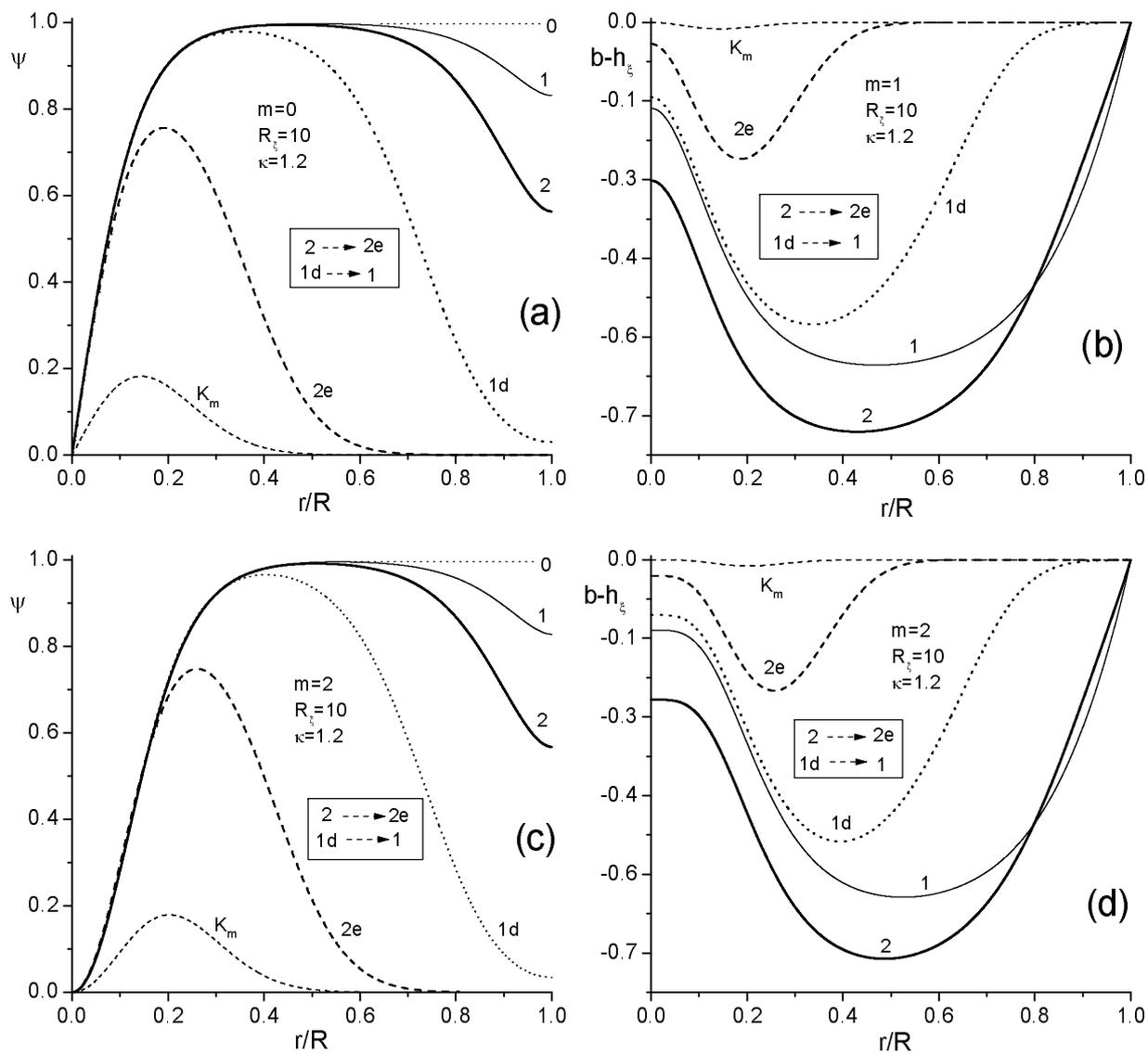
Fig. 4. The coordinate dependences $\psi(x)$ and $b(x)$, as they
are seen through the peep-holes in Fig. 3(a) at $R_\xi=10$,
$\kappa=1.2$ ($m=1$), and in Fig. 3(b) ($m=2$). The curves in (a)
and (b) are calculated for the fields $h_\xi$: {\it 0} --
$h_\xi=0$; {\it 1} -- $h_\xi=0.6367$; {\it 1d} -- $h_\xi=0.6368$;
{\it 2} -- $h_\xi=0.7610$; {\it 2e} -- $h_\xi=0.7611$; $K_m$ --
$h_\xi=0.986$. The jump at the transition {\it 1d}$\to${\it 1} is
$\delta_1=0.018$; the jump at the restoration transition {\it
2}$\to${\it 2e} is $\Delta_r=0.238$. The curves in (c) and (d) are
calculated for the fields $h_\xi$: {\it 0} -- $h_\xi=0$; {\it 1}
-- $h_\xi=0.6358$; {\it 1d} -- $h_\xi=0.6359$; {\it 2} --
$h_\xi=0.7619$; {\it 2e} -- $h_\xi=0.7620$; $K_m$ --
$h_\xi=0.986$. The jump at the transition {\it 1d}$\to${\it 1} is
$\delta_1=0.030$; the jump at the restoration transition {\it
2}$\to${\it 2e} is $\delta_r=0.244$.

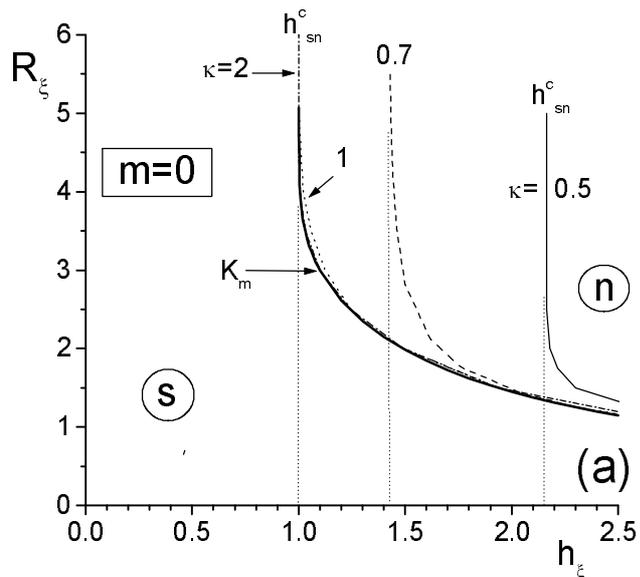
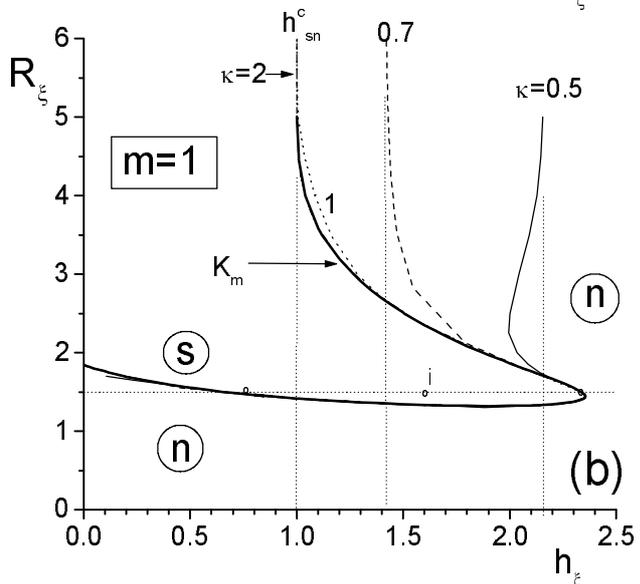
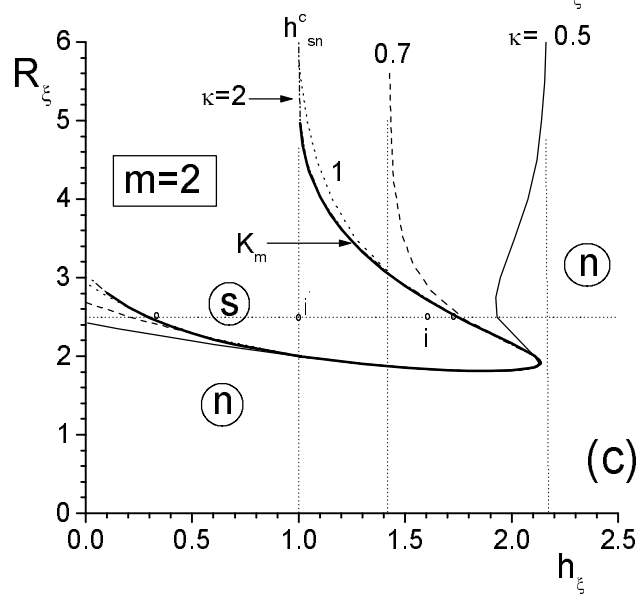
Fig. 5. The phase diagrams represent the critical fields
$h_\xi=H/H_\xi$, at which the transition from $s$- to $n$-state
occurs for a given $R_\xi=R/\xi$ and various $\kappa$ (see the
numerals at the curves) and different vorticities $m$: (a) --
$m=0$; (b) -- $m=1$; (c) -- $m=2$. Thick curves $K_m$ correspond
to the solutions of linear equation (7). The curves with
$\kappa>1$ are well represented by $K_m$-curve [thus, the linear
theory [5] gives correct description of ($s,n$)-boundary].
However, for $\kappa<1$ (and large $R$) the curves deviate
strongly from the curve $K_m$ [thus, for $\kappa<1$ the linear
theory [5] is not applicable]. For sufficiently small $R_\xi$
(when ($s,n$)-transition is of second order) the linear theory is
valid for all $\kappa$.

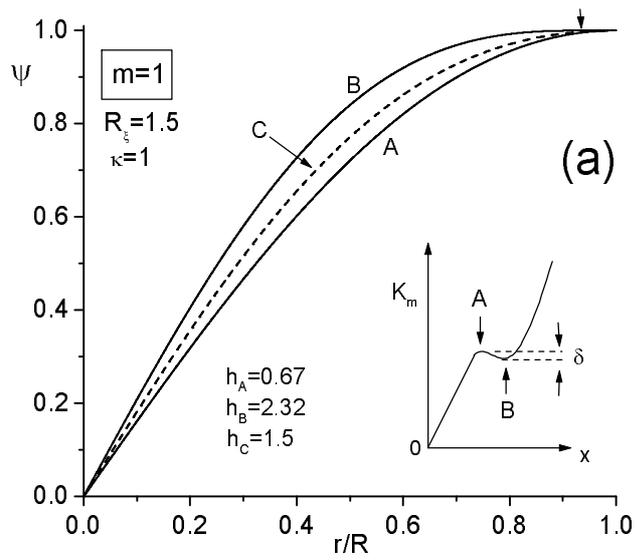
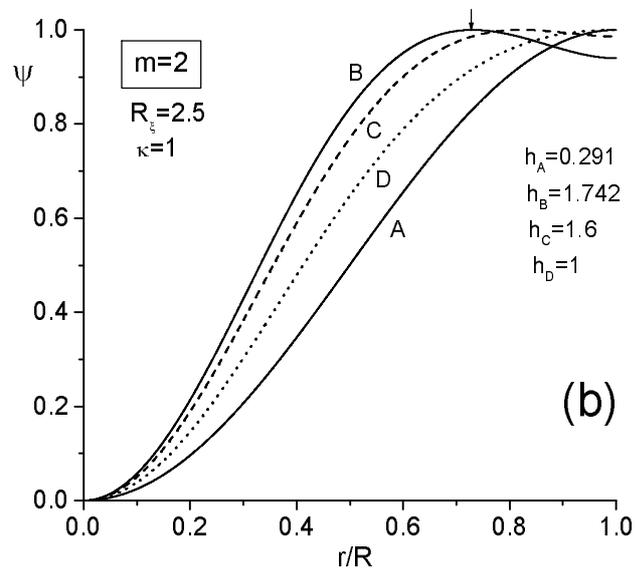
Fig. 6. Two solutions ({\it 1} and {\it 2} ) of the linear
equation (7) in the case: (a) -- $m=1$ [$R_\xi=1.5$, the fields
$h_\xi=0.67$ ({\it 1} ) and $h_\xi=2.32$ ({\it 2} )]; (b) -- $m=2$
[$R_\xi=2.5$, the fields $h_\xi=0.291$ ({\it 1} ) and
$h_\xi=1.742$ ({\it 2} )]. The insert to Fig. 6(a) shows the
schematic behavior of two possible solutions. The zigzag amplitude
$\delta$ for the solution {\it 2} is not seen in this scale in
(a), but is present in (b). (The arrows at the curves {\it 2} show
the maximums of $\psi(x)$.) The dashed curves {\it 3} represent
solutions of the nonlinear $\widetilde{K}_m$-equation (6) in the
intermediate peep-hole $i$ [with $h_\xi=1.5$ (a) and $h_\xi=1.6$
(b)]. The dotted curve {\it 4} represents the self-consistent
solution, seen through the peep-hole $i'$ ($h_\xi=1$) in Fig.
5(c). [All the curves in Fig. 6 are normalized to $\psi_{max}=1$.]

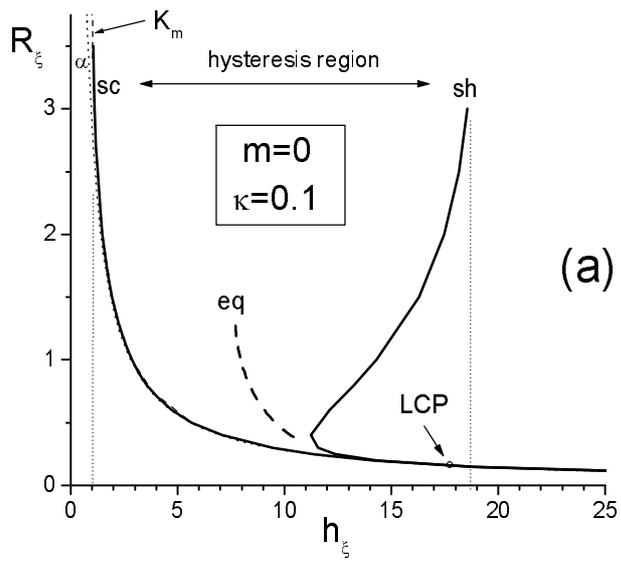
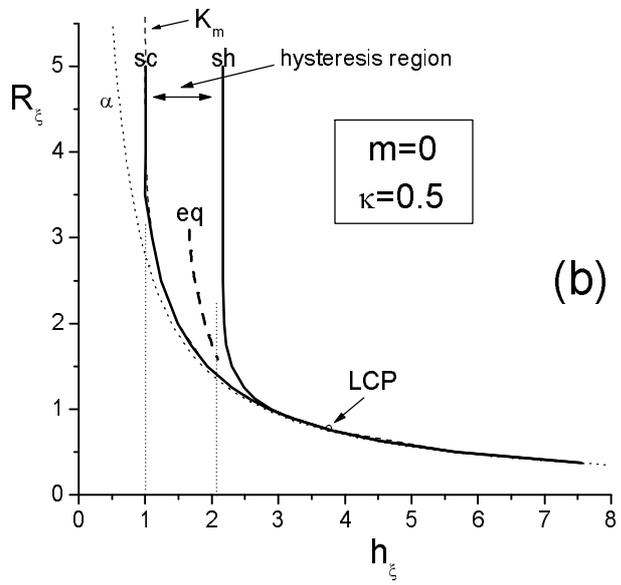
Fig. 7. The upper critical field (the curve $sh$) for superheated
$s$-state, and the lower critical field (the curve $sc$) for
supercooled $n$-state of the vortex-free Meissner state ($m=0$)
for $\kappa=0.1$ (a) and $\kappa=0.5$ (b). For $\kappa=1$ the
$sc$- and $sh$-curves practically coinside. The dotted curve
$\alpha\sim 2.8/h_\xi$ gives good approximation to $sc$-curve for
$R_\xi<2$. The broken curve $eq$ (schematic) corresponds to the
equilibrium transition between $s$- and $n$-states. LCP is the
tri-critical Landau point. The curve $sc$ coincides with the phase
boundary $K_m$ (dashed line), found from the $\kappa$-independent
linear equation (7).

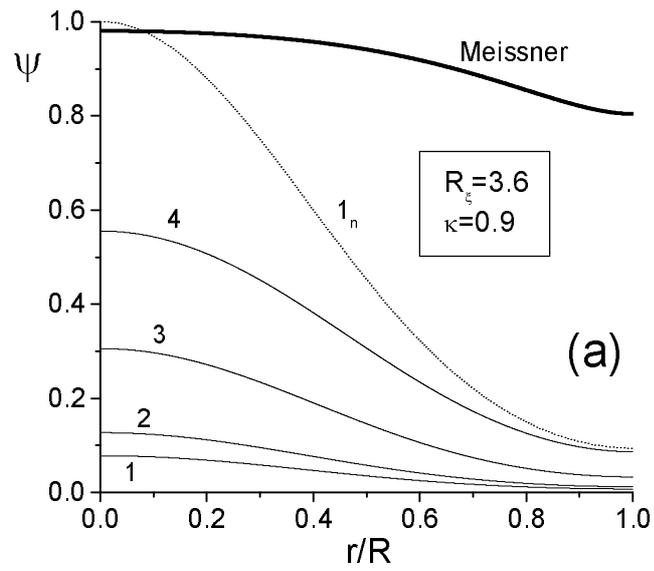
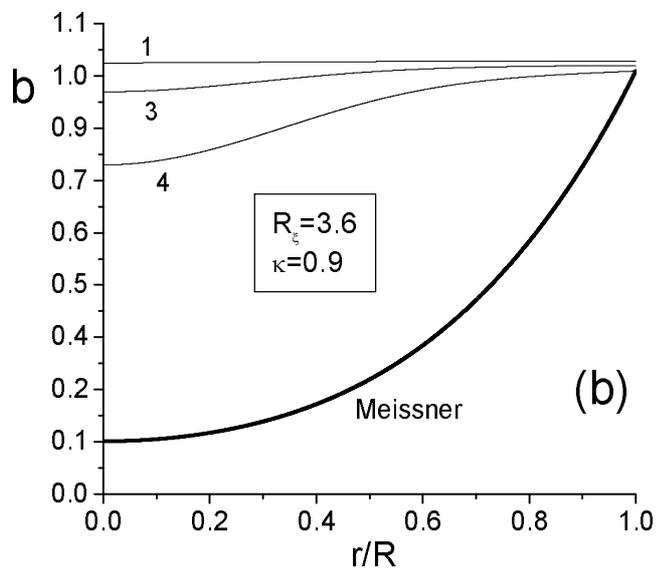
Fig. 8.  Precursor solutions for $R_\xi=3.6$, $\kappa=0.9$: (a) --
$\psi(x)$, (b) -- $b(x)$. Supercooled precursor state nucleates at
$h_p=1.0231$ (with $\psi\approx 0$), and takes forms: {\it 1} --
at $h_\xi=1.0222$; {\it 2} -- at $h_\xi=1.0210$; {\it 3} -- at
$h_\xi=1.0123$; {\it 4} -- at $h_\xi=0.9991$ (this is maximally
supercooled precursor state, which at $h_r=0.9990$ passes into the
Meissner form). The dotted line {\it 1}${}_n$ is the normalized
(Kummer) function {\it 1} ($\psi\ll 1$), which can not be
described as $\psi(x)\approx {\rm const}$ [3]. However, the exact
solution {\it 1} is small ($\psi(x)\approx 0={\rm const}$), so the
field of maximal supercooling of $n$-state  ($h_p$) may be found
from the Ginzburg [3] approximation (and also from the linear
theory [5]).

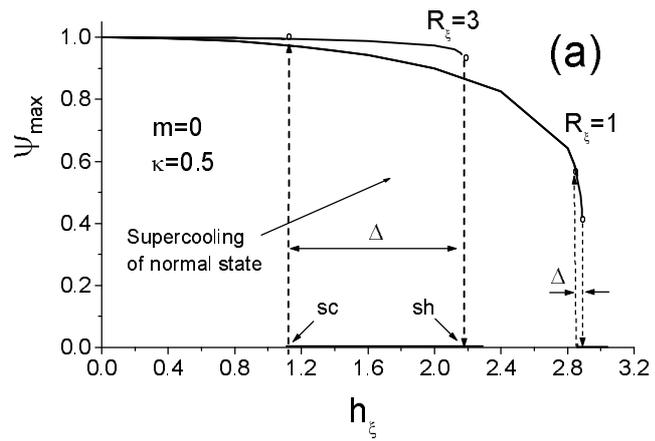
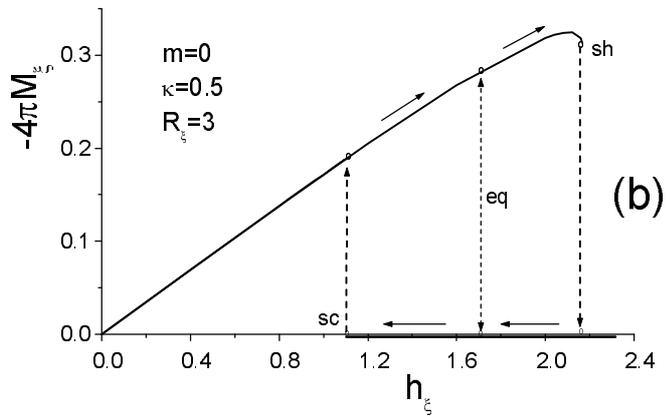
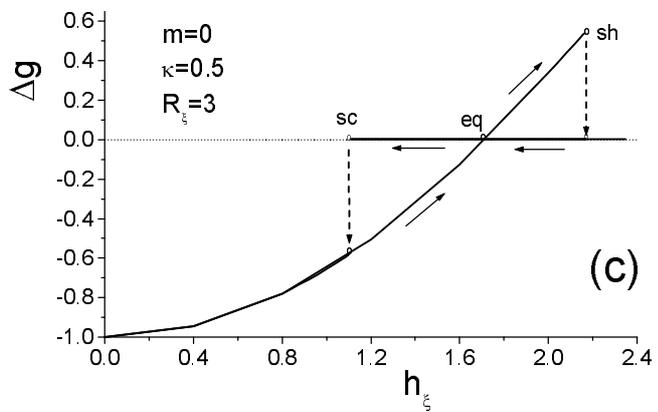
Fig. 9. The field dependence of the order parameter $\psi_{max}$
(a), the magnetization ($M_\xi=M/H\xi$) (b) and the free energy
$\Delta g$ (c) in the Meissner state ($m=0$) of the cylinder with
$\kappa=0.5$ and $R_\xi=3$. The curve $R_\xi=1$ in (a)
demonstrates, that the hysteresis region $\Delta$ vanishes for
small $R_\xi<1$.

Fig. 10. (a) -- Free energy functional $\Delta G$ versus
$\psi={\rm const}$, according to the Ginzburg approach [3]
(schematic). The sequence of curves {\it 1--5} corresponds to the
field $H$ increasing. Minimums of $\Delta G$ correspond to stable
states, maximums -- to unstable states. (b) -- The order parameter
$\psi_0$, found from the extremums of $\Delta G$. Stable branch is
depicted by solid line, unstable -- by dashed line. The numerals
{\it 1--5} correspond to those in (a). The unstable branch does
not exist in the self-consistent theory (see Fig. 9(a)).

\end{document}